\documentclass[10pt,twocolumn]{article}
\usepackage{amsmath}
\usepackage{amsfonts}
\usepackage{amssymb}
\usepackage{graphicx}
\usepackage{bm}
\title{\textbf{Dissociation dynamics in low energy electron attachment to nitrogen dioxide}}
\author{\textbf{Anirban Paul$^{1\dag}$}, \textbf{Dipayan Biswas$^{2}$} and \textbf{Dhananjay Nandi$^{1,3*}$}\\ $^1$ Indian Institute of Science Education and Research Kolkata, Mohanpur 741246, India\\ $^2$ Department of Physics, University of Arizona, Tucson, Arizona, 85721, USA\\ $^3$ Center for Atomic, Molecular and Optical Sciences $\&$ Technologies,\\ Joint initiative of IIT Tirupati $\&$ IISER Tirupati, Yerpedu, 517619, Andhra Pradesh, India \\ \small{email: $^\dag$anirban.paul1995@gmail.com,$^*$dhananjay@iiserkol.ac.in}}

\date{}
\begin{document}
\twocolumn[
  \begin{@twocolumnfalse}
    \maketitle
    \begin{abstract}
    Complete dissociation dynamics of low energy electron attachment to nitrogen dioxide around 8.5 eV resonance has been studied using a velocity map imaging (VMI) spectrometer.
Besides the three prominent resonant peaks at around 1.4 eV, 3.1 eV, and 8.5 eV, we have found an additional small resonance at the higher energy tail of the 8.5 eV resonance.
We have collected the momentum distribution data of O$^-$ ions at different incident electron energies around the 8.5 eV resonance along with the smaller additional resonant peak. 
A theoretical analysis of these resonances with the momentum imaging experimental data on dissociative electron attachment to nitrogen dioxide in the gas phase is used to provide a detailed picture of the molecular dissociation process.
    \end{abstract}
  \end{@twocolumnfalse}
]

\section{Introduction}

Dissociation of molecules induced by the collision of electron-like charged particles is a vital process in different branches of science since it is directly related to the mechanisms of radiation-induced damage of living cells and depletion of the ozone layer in the upper atmosphere. Damage to living cells due to nuclear radiation is mainly caused by dissociative electron attachment (DEA) by the low-energy secondary electrons generated from higher-energy primary radiation. Single and double-strand breaks of DNA \cite{Boudaiffa} are also primarily caused by low-energy secondary electron impact. Dissociative electron attachment (DEA), one of the most important mechanisms in these physical processes, is a two-step resonant process dominant in low-energy electron-molecule inelastic collisions. In the first step, the incident electron is attached to the molecule, forming a temporary negative ion (TNI) state. In most cases, this negative ion state is very unstable and repulsive. This unstable TNI dissociates into a negative ion fragment and one or more neutral fragment(s) in the subsequent step.
It is, therefore, crucial to perform similar experiments in the laboratory in a controlled atmosphere \cite{Kawarai, Mahmoodi-Darian, Sahbani}.\\
Nitrogen dioxide(NO$_2$) is an atmospheric gas that can act as a free radical with an unpaired electron in its highest occupied molecular orbital (HOMO). It exists in equilibrium with its dimer N$_2$O$_4$. It is a toxic gas and an industrial pollutant responsible for photochemical smog. It can cause the depletion of the ozone layer and play a significant role in the decomposition of the ozone layer. In the air, toxic nitric oxide (NO) and other organic nitrates are also formed from NO$_2$. It is a bent-shaped molecule having a C$_{2v}$ point symmetry which gives some interesting complexity for its dissociation, which makes it much more interesting to study from the purely scientific point of view.\\
DEA to NO$_2$ has been studied earlier by many groups. Most of them have measured the appearance energy, peak positions of the resonances, and relative cross-section of these resonances. 
Fox \cite{Fox} observed O$^-$ ion peaks at 1.9 eV, 3.0 eV, and 8.75 eV with onset values at 1.35 eV, 2.5 eV, and 7.3 eV and concluded that the peak around 8.75 eV is due to the impurities like NO or H$_2$O in NO$_2$. 
Rallis and Goodings \cite{Rallis} observed O$^-$ ion peaks at 3.0 eV and 8.1 eV with their onset values at 1.6 eV and 7.3 eV. They found the first peak onset matched well with the thermodynamical threshold value 1.65 eV and concluded that the 3 eV resonant peak, the O$^-$ ions are produced with NO in its electronic ground state. While for the resonance peak at 8.1 eV, the O$^-$ ions are created with NO in its first electronic excited state $( a^4 \Pi )$. 
Abouaf and Fiquet- Fayard \cite{Abouaf} found that the O$^-$ ions in the first resonant peak are produced with NO $( X^2 \Pi )$ via the dissociation of NO$_2$ in the $^1$B$_1$ resonance state.
Rangwala \textit{et al}. \cite{Rangwala} measured the absolute DEA cross-sections of O$^-$ ions from NO$_2$ and observed the peak position of the resonances at 1.4 eV, 3.1 eV, and 8.3 eV. Based on the R-Matrix calculations, Munjal \textit{et al.} showed that NO$_2^-$ supports a bound state A below the ground state of NO$_2$.\cite{Munjal} They found two shape resonances $^3$B$_1$ and $^1$B$_1$ at 1.18 and 2.3 eV respectively are responsible for the first two 1.8 and 3.1 eV peaks.
Later Gupta \textit{et al.} calculated the total scattering cross-section for NO$_2$ + e$^-$ and compared it with the absolute scattering cross-sections measured by Szmytkwoski \textit{et al.} \cite{Gupta} They also obtained the resonance positions in total scattering cross-section at 1.33 and 3 eV having symmetries $^3$B$_1$ and $^1$B$_1$, respectively.
Nandi and Krishnakumar \cite{Dhananjay} measured the kinetic energy distribution of fragment O$^-$ ions using the time of flight (TOF) mass spectrometer technique around those resonances. They found that for the 1.8 eV and 3.5 eV resonant peaks, the kinetic energy vs. incident electron energy curves have a small slope, and their threshold values are very close. Therefore, the excess energy is distributed as the internal energy of the NO fragments, and both of these resonances have the same dissociation limit. For 8.5 eV resonance, they assigned the dissociation channel leads to O$^-$ ion is  O$^-$  + NO $(a^4 \Pi)$ with a threshold energy of 6.43 eV.
Gope et al. \cite{Gope} measured the kinetic energy and angular distribution of  O$^-$ ions at different incident electron energies around the third resonance peak, which is at 8.5 eV energy using Velocity Slice Image (VSI) technique. They found both the low and high energy O$^-$ ions at this resonance and attributed the low energy ions to O$^-$  + NO $(A^2 \Sigma^+)$ channel having threshold energy 7.10 eV, while the high energy ions to O$^-$  + NO $(D^2 \Sigma^+)$ channel having threshold energy 8.20 eV. They fitted the angular distribution of the high energy ions with different resonant symmetries and their combination and found that the best fit is for B$_1$ + B$_2$ resonant symmetry. In contrast, the previous theoretical studies found no B$_1$ resonance in this range.
Tian and co-workers recently studied the complete dissociation dynamics of dissociative electron attachment to NO$_2$ at the two low-energy resonances.\cite{Tian2023}
They found the involvement of a $^3$B$_2$ resonant state in the 3.1 eV resonance.\cite{Tian2023}\\
In this article, we have reported the detailed systematic studies of DEA to NO$_2$ around the higher energy resonances using the velocity slice imaging (VSI) technique.\cite{rsi:DN,dipayan:co,Prabhudesai_Acetone,cs2,o2:pamir} Here, we report distinct kinetic energy and angular distributions of the fragment ions. Our angular distribution data matches quite well with the theoretical prediction.

\section{Experimental}
All the experiments have been carried out in our velocity slice imaging (VSI) setup. The details of this setup have been reported earlier \cite{nag2015fragmentation, Nag_2015} elsewhere. Therefore, we have briefly discussed the setup and experimental procedure here.
The setup mainly consists of an electron gun, a Faraday cup, and a velocity slice imaging (VSI) spectrometer. The electron gun produces a pulsed (200 ns wide) electron beam which is further collimated magnetically by the magnetic field produced by two coils in Helmholtz configuration.
This collimated electron beam is made to cross at right angles with an effusive molecular beam produced by a capillary along the axis of the Velocity Slice Imaging (VSI) spectrometer. The spectrometer consists of a single electrostatic lens and a conical drift tube. 
The VSI spectrometer is followed by a 2D position-sensitive detector (PSD) consisting of three microchannel plates mounted in a Z-stack configuration and a hexanode position-sensitive detector placed just after the MCP stack. A pulsed electric field extracts the newton sphere of the ions formed in the interaction region applied 100 ns after the electron beam pulse. 
The main principle \cite{Eppink} of the VMI spectrometer is to focus all the ions having the same velocity (speed + direction) on a single point to the detector. The lens plate is used for this focusing. After the lens plate, a conical-shaped drift tube is used; for our experimental purpose, we have applied 108 V potential to the drift tube. The drift tube is used to expand the newton sphere of the fragment ions, and we get better resolution. After that, there is a two-dimensional position-sensitive detector consisting of three microchannel plates (MCPs) in a Z-stack configuration. Behind these MCPs is a delay line hexanode \cite{Jagutzki}. The time-of-flight (TOF) of the detected ions is determined from the back MCP signal, and hexanode plates collect the x and y positions of each detected ion. Thus, each ion's x, y, and ToF can be collected and stored in a list-mode file format (.lmf); for our experimental purpose, we have used the CoboldPC software from RoentDek to collect these data and to store them in a .lmf file. Thus, we can construct the full newton sphere of the negative ions from the .lmf file. 
\begin{figure}[h]
\centering
  \includegraphics[scale=0.36]{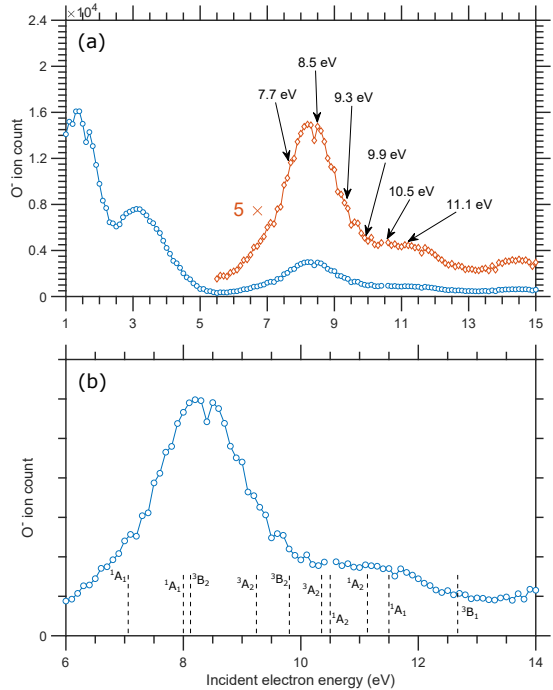}
  \caption{\footnotesize{(a) Ion-yield of O$^-$ ions produced due to DEA to NO$_2$. The black arrows indicate the energies at which the velocity slice images are taken. (b) O$^-$ yield for the incident electron energy in the range of 6 to 14 eV. The dashed vertical lines indicate the energy of different resonant states found by our theoretical calculation.}}
  \label{fig:ion-yield}
\end{figure}\\
For our analysis, we have used solid angle weighted slices \cite{Moradmand2013,cs2}. We used the data of DEA to Oxygen (O$_2$) for calibration. We have performed the experiments using  99.9 $\%$ pure commercially available NO$_2$ gas. 

\section{Computation}
We conducted ab initio electronic structure and fixed-nuclei electron scattering calculations to interpret our results. 
NO$_2$ is a bent triatomic molecule with C$_{2v}$ point group symmetry. It has 1.2 \r A equilibrium bond length, while it has 134.5$^\circ$ bond angle at equilibrium. It is an open-shell molecule nominally described, near its equilibrium geometry, by the electronic configuration. The ground state electron configuration of NO$_2$ is $(1a_1)^2 (1b_2)^2 (2a_1)^2 (3a_1)^2 (2b_2)^2 (4a_1)^2 (5a_1)^2 (3b_2)^2$ $(1b_1)^2 (4b_2)^2 $ $ (1a_2)^2 (6a_1)^1$ and its ground state is $X^2A_1$. 
The neutral and anion states are described by multi-reference configuration-interaction (MRCI) calculations that include single- and double-excitations from a complete active space (CAS). The CAS orbitals are obtained from state-averaged multi-configuration-self-consistent-field (MCSCF) calculations. 
The ground energy of the neutral molecule is optimized by restricting the first 16 electrons in the first eight orbitals, while the remaining 7 electrons are kept free in the active space of 6 molecular orbitals. The energy we found from this calculation is -204.152 H, which agrees very well with the previously found values -201.144 H by Munjal \textit{et al.} and -204.15 H by Gupta \textit{et al.} The obtained dipole moment is 0.374 D, which is in fair agreement with the experimental value of 0.316 D.\cite{Leonardi}
On the other hand, for anions, out of 24 electrons, we froze 12 in the first six molecular orbitals to determine the energy of the anion states, while the remaining 12 electrons were kept free in the active space of 10 molecular orbitals for the MCSCF calculation. 
\begin{figure*}[h]
 \centering
    \centerline{\includegraphics[scale=0.4]{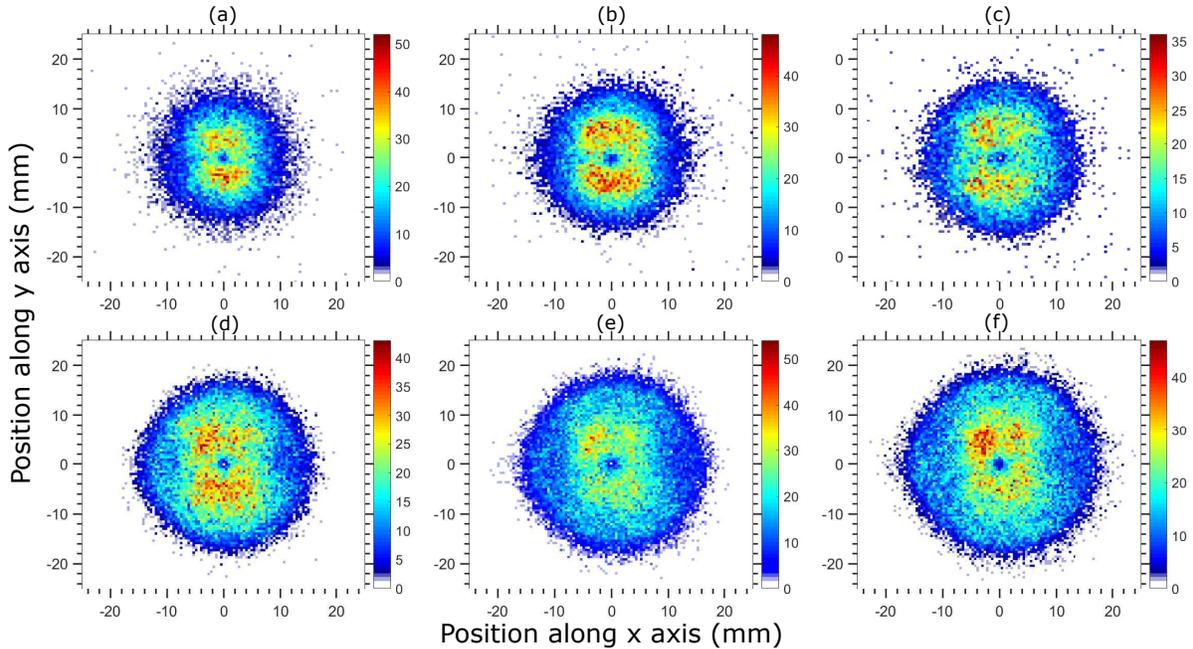}}
  \caption{\footnotesize{ Wedge sliced images of O$^-$ /NO$_2$ ions for (a) 7.7 eV (b) 8.5 eV, (c) 9.3 eV, (d) 9.9 eV, (e) 10.5 eV, and (f) 11.1 eV incident electron energy.
  The electron beam direction is from left to right (The red arrow indicates the electron beam direction) through the center of each image. The wedge angle used for our analysis is 8$^\circ$ for each of the images.}}
  \label{fig:vsi-images}
\end{figure*}
We performed all these calculations with the program GAMESS-US package. \cite{GAMESS}
As mentioned earlier, we have calculated the energies of different excited states of NO$_2^-$ under the equilibrium geometry condition of the neutral NO$_2$ molecule. According to the Born-Oppenheimer approximation, the electronic motion is much faster than that of a molecule's nuclear motion, and the equation of motions for electrons can be separated from the nuclear motions. Based on this, the Franck-Condon transition principle suggests that the electronic transition occurs under the same geometry as it was initially. Therefore, the resonance energy can be calculated by the energy difference of an anionic state calculated under the neutral equilibrium geometry to that of the neutral molecule.
From this calculation, we get that a $^3$B$_1$ and a $^1$B$_1$ resonances are at 1.02 and 2.02 eV energy. This matches with the result found by Munjal \textit{et al.} who found these at 1.18 and 2.3 eV, respectively.\cite{Munjal} 
\begin{figure}[h]
\centering
  \includegraphics[scale=0.44]{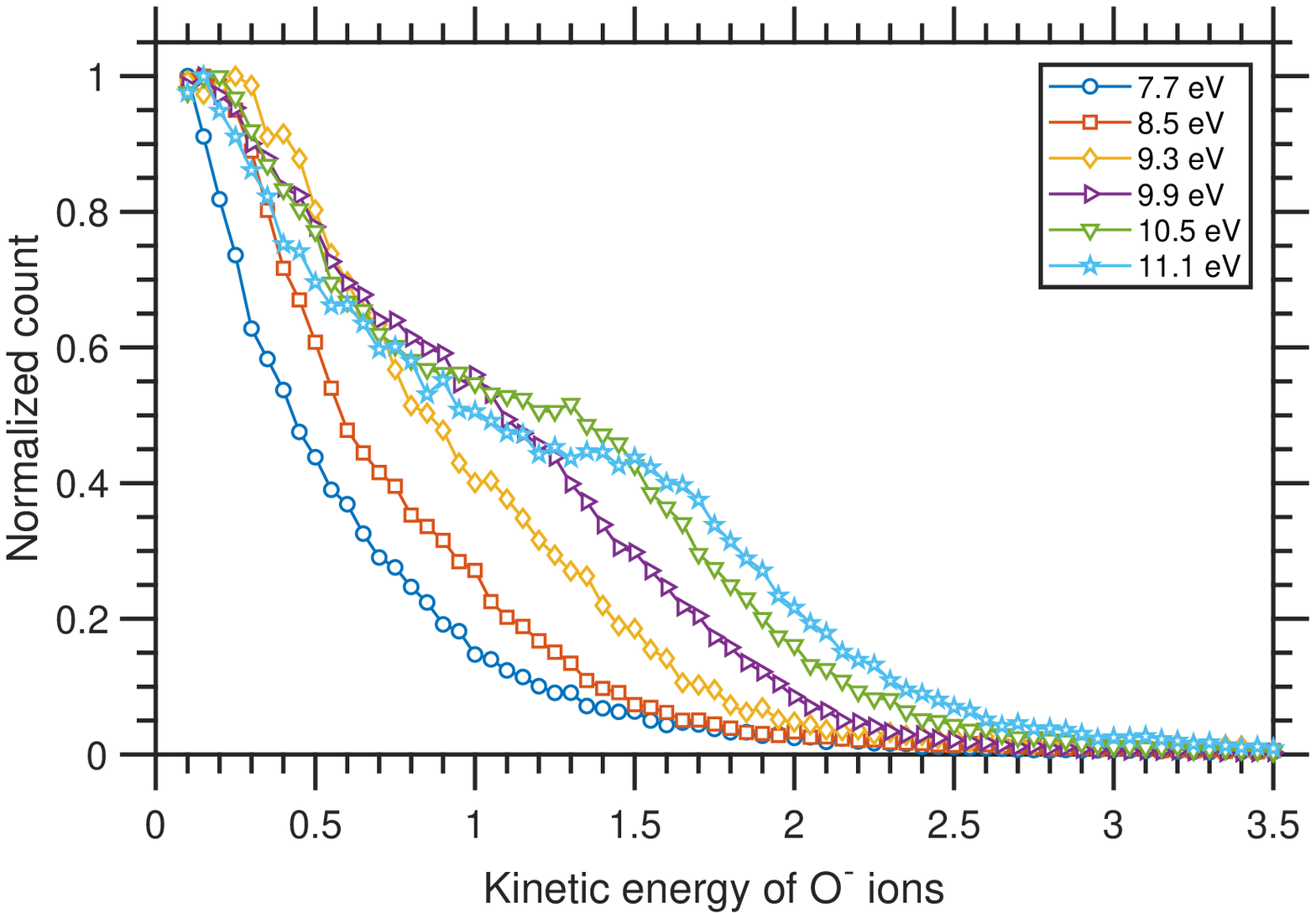}
  \caption{\footnotesize{Kinetic energy distributions of O$^-$ from
DEA to NO$_2$ at different incident electron energies around the 8.5 eV resonance. The distributions shown are obtained after integrating over entire $2\pi$ angles about the electron beam direction and normalized by the peak.}}
  \label{fig:KE_dist}
\end{figure}

\section{Results and Discussion}
In Fig. \ref{fig:ion-yield}, we have plotted the O$^-$ ion yield as the function of electron energy for the 1.0 to 15.0 eV energy range.
The ion yield shows three main resonances, the most intense one is at around 1.4 eV, the second one is at about 3.1 eV, and the third one is at about 8.5 eV. The brown curve in Fig. \ref{fig:ion-yield} is the zoomed view of the excitation function for the 6 to 15 eV range. 
This zoomed view suggests that the third resonance is a combination of two resonances; the bigger one is peaking at around 8.5 eV, while the smaller one is at the falling edge of the bigger one.
The theoretically calculated value of the ion-pair dissociation (IPD) (NO$^+$ + O$^-$) threshold for NO$_2$ is 10.91 eV. \cite{ChakrabortyIPD} 
Upon close inspection of the ion yield curve, an overlap between the DEA and ion-pair dissociation can be observed. Previous studies have evident this kind of overlap between the IPD and DEA.\cite{ChakrabortyCH2F2}
In the present study, we have found another small resonance at the higher energy tail of the main resonance. 
\begin{table}[h]
\centering
\caption{Resonance position, their nature, and symmetry responsible for the broad resonance peaking at 8.3 eV and the smaller resonance at the higher energy tail of that.}
\begin{tabular}{c|c|c}
\hline
Position of the & Symmetry & Electronic\\
resonance (eV) & & configuration \\
\hline
& & \\
7.04 & $^1A_1$ &  1b$_1^2$4b$_2^2$1a$_2^2$6a$_1^0$2b$_1^2$ \\
& & \\
7.99 & $^1A_1$ & 1b$_1^2$4b$_2^2$1a$_2^2$6a$_1^0$7a$_1^2$  \\
& & \\
8.15 & $^3B_2$ & 1b$_1^2$4b$_2^1$1a$_2^2$6a$_1^1$2b$_1^2$ \\
& & \\
9.24 & $^3A_2$ & 1b$_1^2$4b$_2^2$1a$_2^1$6a$_1^1$2b$_1^2$  \\
& & \\
9.78 & $^3B_2$ & 1b$_1^2$4b$_2^1$1a$_2^2$6a$_1^1$7a$_1^2$  \\
& & \\
10.35 & $^3A_2$ & 1b$_1^2$4b$_2^2$1a$_2^1$6a$_1^1$7a$_1^2$  \\
& & \\
10.50 & $^1A_2$ & 1b$_1^2$4b$_2^2$1a$_2^1$6a$_1^1$2a$_1^2$ \\
& & \\
11.14 & $^1A_2$ & 1b$_1^2$4b$_2^2$1a$_2^1$6a$_1^1$7a$_1^2$  \\
& & \\
11.36 & $^1B_2$ & 1b$_1^2$4b$_2^1$1a$_2^2$6a$_1^1$2a$_1^2$  \\
& & \\
11.50 & $^1A_1$ & 1b$_1^2$4b$_2^2$1a$_2^2$6a$_1^0$7a$_1^2$  \\
& & \\
12.66 & $^3B_1$ & 1b$_1^1$4b$_2^2$1a$_2^2$6a$_1^1$2b$_1^2$  \\
& & \\
\hline
\end{tabular}
\label{table:resonances-higher}
\end{table}
We have taken the VSI data of O$^-$ ions for six different incident electron energies around these two resonances. The solid angle weighted velocity slice images of O$^-$ ions are shown in Fig. \ref{fig:vsi-images}.\\
In the present theoretical calculation, we have found several Feshbach resonances present in the FC transition region's 6 to 12 eV energy. In Table \ref{table:resonances-higher}, we have listed the position of the resonances in the FC region. From that table, we see three resonances with A$_1$, three resonances with B$_2$, and four resonances with A$_2$ symmetry are present in that energy range. 
The $^1$A$_1$ resonance at around 7.04 eV is a Feshbach one where one electron from 2$\pi$ orbital gets excited into the 2b$_1$ molecular orbital from 6a$_1$ orbital, and the incident electron also gets caught into the b$_1$ orbital. The electron configuration is also listed in Table \ref{table:resonances-higher}.
The other two A$_1$ resonances (at 7.99 and 11.5 eV) are due to the excitation into the 7a$_1$ and 8a$_1$ orbital respectively from 6a$_1$ orbital.
The electron configuration of each of the resonances has listed in Table \ref{table:resonances-higher}.
From Fig. \ref{fig:ion-yield} (b), we can see that the bigger resonant peak is extended up to about 10 eV. The two A$_1$, one A$_2$, and two B$_2$ resonant symmetric states are present in this energy range. On the other hand, the smaller peak is extended up to about 12.5 eV energy. Three A$_2$, one A$_1$, and one B$_2$ resonant symmetric states are present in that energy range.

\subsection{Kinetic energy distribution}
The kinetic energy distribution of the O$^-$ ions for different incident electron energies extracted for the entire 0 to 2$\pi$ angle about the electron beam direction is shown in Fig. \ref{fig:KE_dist}. 
The kinetic energy distributions in Fig. \ref{fig:KE_dist} show only one near-zero energy peak up to 8.5 eV.
One additional overlapping peak at the higher energy region appears in the distribution for electron energy, $\geq $ 9.3 eV one higher energy peak appears in the distribution.
\begin{table}[h]
\centering
\begin{tabular}{|c|c|c|}
\hline
Chanel & DEA channels &	Threshold \\
no. & & energy (eV)\\
\hline
 1. & O$^-$  + NO $(X^2 \Pi)$ & 1.65 eV\\
\hline
 2. & O$^-$  + NO $(a^4 \Pi)$ & 6.43 eV\\
\hline
 3. & O$^-$  + NO $(A^2 \Sigma^+)$ & 7.10 eV \\
\hline
 4. & O$^-$  + NO $(B^2 \Pi)$ & 7.46 eV \\
\hline
 5. & O$^-$  + NO $(C^2 \Pi)$ & 8.11 eV \\
\hline
 6. & O$^-$  + NO $(D^2 \Sigma^+)$ & 8.20 eV \\
\hline
 7. & O$^-$  + N $(^4S)$ + O $(^3P)$ & 8.18 eV \\ 
\hline
\end{tabular}
\caption{DEA channels of NO$_2$ and their threshold values. \cite{Gope}}
\label{table:1}
\end{table}\\
Gope \textit{et al.} also found both the higher and lower energy ions at this resonance.\cite{Gope} From their threshold calculation, they found that the low energy ions are produced from  O$^-$  + NO $(A^2 \Sigma^+)$ dissociation channel having threshold energy 7.10 eV. In contrast, the higher energy ions are produced from  O$^-$  + NO $(D^2 \Sigma^+)$ dissociation channel having a threshold energy of 8.20 eV.

\subsection{Angular distribution}
\begin{figure*}[h]
 \centering
    \centerline{\includegraphics[scale=0.24]{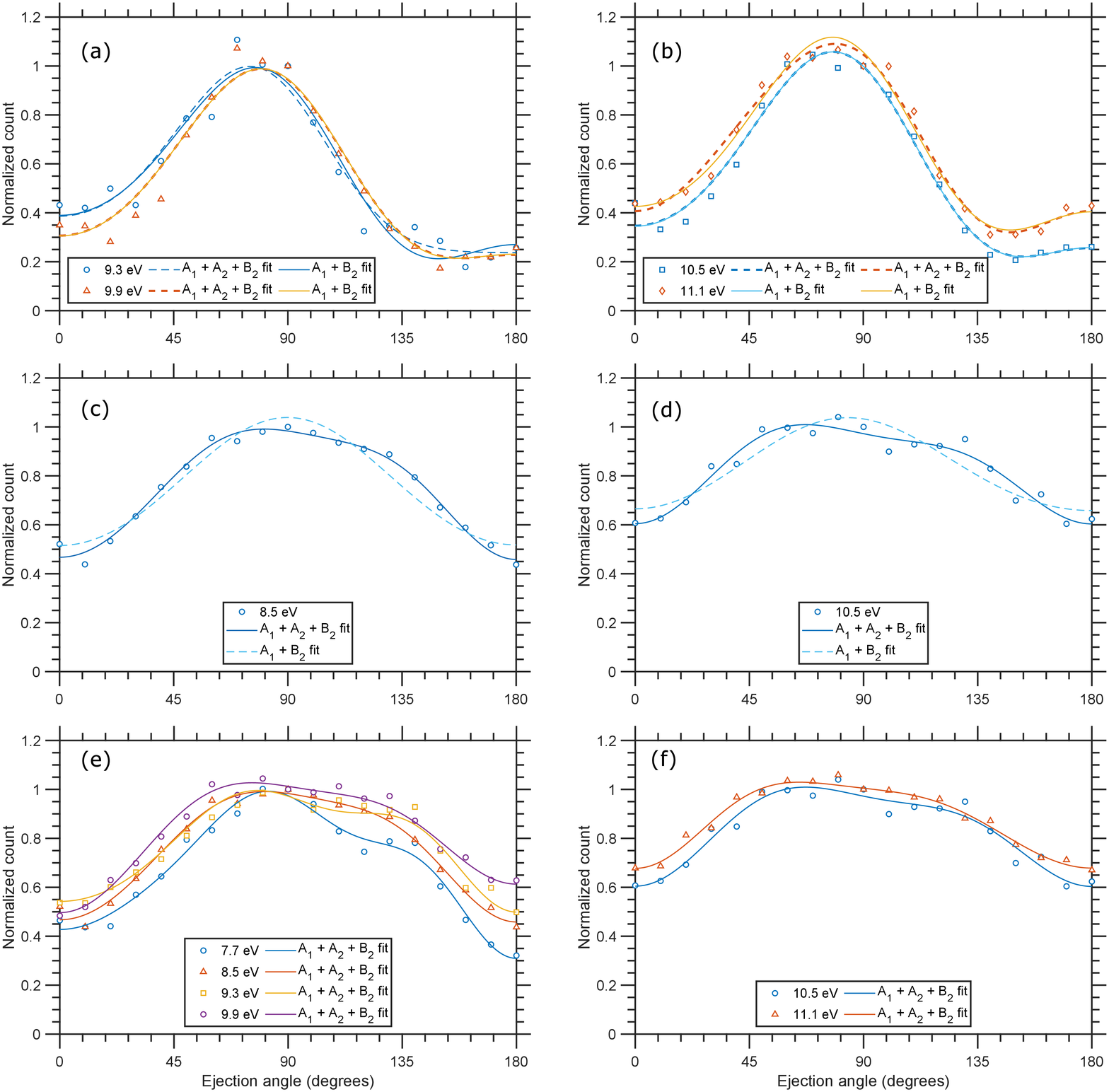}}
  \caption{\footnotesize{(a) Data points represent the experimentally obtained angular distributions of the O$^-$ ions in the higher energy band for 9.3 eV (Blue) and 9.9 eV (Orange) incident electron energy, respectively. The solid curves represent the best fits with A$_1$+B$_2$ resonant symmetries. While the dashed curves represent the best fits with A$_1$+A$_2$+B$_2$ resonant symmetries. (b) Data points represent the experimentally obtained angular distributions of the O$^-$ ions in the higher energy band for 10.5 eV (Blue) and 11.1 eV (Orange) incident electron energy, respectively. The solid curves represent the best fits with A$_1$+B$_2$ resonant symmetries. In comparison, the dashed curves represent the best fits with A$_1$+A$_2$+B$_2$ resonant symmetries.
 (c) Data points represent the experimentally obtained angular distributions of the O$^-$ ions in the higher energy band for 8.5 eV incident electron energy. The black solid curve represents the best fit with A$_1$+B$_2$ resonant symmetries. While the solid blue curve represents the best fit with A$_1$+A$_2$+B$_2$ resonant symmetry. (d) Angular distributions of the O$^-$ ions in the lower energy band for 10.5 eV incident electron energy. (e) Angular distributions O$^-$ ions in the lower energy band around the bigger resonant peak fitted with A$_1$+A$_2$+B$_2$ resonant symmetries. (f) Angular distributions O$^-$ ions in the lower energy band around the smaller resonant peak fitted with A$_1$+A$_2$+B$_2$ resonant symmetries. }}
  \label{fig:AD_All}
\end{figure*}

The angular distributions of the fragment O$^-$ ions have been extracted from the sliced images.
We have plotted the angular distribution of the lower energy ions for the 7.7, 8.5, 9.3, 9.9, 10.5, and 11.1 eV incident electron energies in fig. \ref{fig:AD_All} (a). For the low-energy ions, we have considered the ions with kinetic energy $\leq$ 1.2 eV.
While the angular distribution of the higher energy ions for the 9.3, 9.9, 10.5, and 11.1 eV incident electron energies in fig. \ref{fig:AD_All} (b). For the higher-energy ions, we have considered the ions with kinetic energy $\geq$ 1.4 eV.
We can see from fig. \ref{fig:AD_All} (a) that the angular distribution of the low energy ions has a slight dip at around 80$^\circ$ angle with a relatively higher backward count. The angular distribution of the low-energy ions remains more or less similar with the increase in the incident electron energy. With the incident electron energy increase, the backward count increases, and the dip also becomes prominent.
On the other hand, from the angular distributions of the higher energy O$^-$ ions [Fig. \ref{fig:AD_All} (b)] we can see that the ion count has a peak around 80$^\circ$ and also have some finite count in forward and backward direction with a small forward-backward asymmetry. For the angular distributions of the higher energy ions, the distribution remains more or less similar with the increase in the incident electron energy. With the increase in the incident electron energy, the backward count decreases, and the dip near 150$^\circ$ becomes prominent. Our angular distribution agrees reasonably well with the distribution found by Gope \textit{et al.}
They also found the distribution peaks at around 90$^\circ$ with small contributions in the forward and backward directions. \cite{Gope}\\
As discussed earlier, the $^1$A$_1$ at 7.04 eV, $^1$A$_1$ at 7.99 eV, $^3$B$_2$ at 8.15 eV, $^3$A$_2$ at 9.24 eV, and $^3$A$_2$ at 9.78 eV are the possible resonances for the prominent resonant peak at 8.5 eV.
On the other hand, $^3$A$_2$ at 10.35 eV, $^1$A$_2$ at 10.50 eV, $^1$A$_2$ at 11.14 eV, $^1$B$_2$ at 11.36 eV, and $^1$B$_2$ at 11.50 eV are the possible resonant states for the small resonant peak at the higher energy edge of the bigger one. To verify this, we have also fitted these experimentally observed angular distributions with the theoretically predicted ones.\\
The expression for the angular distribution of the fragment negative ions from the DEA to the diatomic molecule was first given by O'Malley and Taylor \cite{O'Malley}.
The expression is as follows
\begin{equation}
 I(\theta,\phi,k) = \sum_{ \mu } \vert \sum_{l = \mu } a_{l \mu}(k)Y_l^\mu(\theta,\phi)e^{i\delta_l}\vert^2   
\end{equation}
where $a_{l\mu}(k)$ are energy-dependent expansion coefficients, $k$ is the incident electron momentum, $Y_l^{\mu}(\theta,\phi)$ are the spherical harmonics, $\mu$ is the difference in the projection of the angular momentum along the inter-nuclear axis for the neutral molecular state and the negative ion resonance state, given as $\mu = \vert \Lambda_f - \Lambda_i \vert$, $l$ is the angular momentum of the incoming electron with values given by $l \geq \vert \mu \vert$ and ($\theta$, $\phi$) are the polar angles of the negative ion fragments with respect to the incident electron direction.
Later Azaria \textit{et al.} \cite{Azria} extrapolated this for the polyatomic molecules and found the angular distribution of the negative ion fragments averaging over $\phi$. The expression is as follows:
\begin{eqnarray}
I(\theta) = \frac{1}{2 \pi}\int_{ 0 }^{2 \pi} \vert \sum_{l \mu \epsilon } i^l e^{i\delta_l} a_{l\mu}^{\epsilon} X_{l \mu}^{\epsilon}(\theta,\phi)\vert^2 d \phi
\end{eqnarray}
where $X_{l\mu}^{\epsilon}$ are the basis functions for the irreducible representation of the group of the molecule, $a_{l\mu}^{\epsilon}$ are their amplitude and all other variables are the same as discussed earlier.
\begin{table}[h]
\centering
\begin{tabular}{|c|cccc|c|}
\hline
C$_{2v}$ & E & C$_2$ & 
$\sigma_v$ & $\sigma_v'$ & Basis functions\\
\hline
 A$_1$ & +1 & +1 & +1 & +1 & Y$_l^m$+Y$_l^{-m}$ ; m=even\\
&&&&&\\
 A$_2$ &  +1 & +1 & -1 & -1 & Y$_l^m$-Y$_l^{-m}$ ; m=even\\
&&&&&\\
 B$_1$ &  +1 & -1 & +1 & -1 & Y$_l^m$+Y$_l^{-m}$ ; m=odd\\
&&&&&\\
 B$_2$ &  +1 & -1 & -1 & +1 & Y$_l^m$-Y$_l^{-m}$ ; m=odd\\ 
\hline
\end{tabular}
\caption{Character table of C$_{2v}$ point group  and basis functions.}
\label{table:Character_table_C2v}
\end{table}\\
To have all the functions defined in the same coordinates and to be able to compare them with the measurements, we need to transform the basis functions (table \ref{table:Character_table_C2v}), and the expression for the partial waves via the Euler angles to the dissociation frame of the molecule.
The angles we used are (0, $\beta$, 0) for the basis functions, where $\beta$ is the half-bond angle.  For the partial wave representing the electron beam, we used ($\phi$, $\theta$, 0) which are the polar angles of the electron momentum vector in the dissociation frame.
The expression of the angular distribution of the O$^-$ ions for A$_1$ resonant state symmetry under the axial recoil approximation is as follows:
\begin{align}
& I^{A_1}_{s+p+d}(\theta)=a_0^2 + a_1^2 (sin^2\beta \, sin^2\theta + cos^2\beta \, cos^2\theta)+ \nonumber \\
& a_2^2 [\frac{9}{16} (sin^4 \beta sin^4 \theta 
 + sin^2 2\beta \, cos^2 2\theta) + \nonumber \\
&\frac{1}{2} (3 cos^2 \beta - 1)^2 (3 cos^2 \theta - 1)^2]+ \nonumber \\
& 4 a_0 a_1 cos \beta \, cos \theta \, cos \delta_0 + \nonumber \\
& 2 a_1 a_2 [\frac{3}{4} sin \, \beta \, sin \, 2 \beta \, sin \, \theta \, sin\,2 \theta + \nonumber \\
& \frac{1}{2} cos \, \beta \, cos \, \theta \, (3 cos^2 \theta - 1)(3 cos^2 \theta - 1)] \, cos \, \delta_1 + \nonumber \\ 
& a_0 a_2 (3 cos^2 \theta - 1)(3 cos^2 \theta - 1) cos (\delta_0 + \delta_1)
\end{align}
Here, we have considered up to the d-partial wave (l=2).
The expression of the angular distribution for other resonant symmetries is as follows:
\begin{align}
I^{A_2}_d(\theta) = & cos^2 \beta \, sin^4 \theta + sin^2 \beta \, sin^2 2\theta \\
I^{B_1}_{p+d}(\theta) = & 2 b_1^2 sin^2\theta + \frac{3}{2} b_2^2 [sin^2 \beta sin^4 \theta + cos^2 \beta sin^2 2\theta]  \nonumber \\
& + 2 b_1 b_2 \sqrt{3} cos \, \beta \, sin \, \theta \, sin\,2 \theta \, cos \, \delta_2 \\
I^{B_2}_{p+d}(\theta) = & 2 b_3^2 (sin^2\beta \, cos^2\theta + cos^2\beta \, sin^2\theta)+ \nonumber \\
& \frac{3}{2} b_4^2 [\frac{1}{4} sin^2 2\beta sin^4 \theta + \nonumber \\
& cos^2 2\beta sin^2 2\theta + \frac{1}{2} sin^2 2\beta (3 cos^2 \theta - 1)^2]+ \nonumber \\
& 2 b_3 b_4 [\sqrt{3} cos \, \beta \, cos \, 2 \beta \, sin \, \theta \, sin\,2 \theta + \nonumber \\
& sin \, \beta \, sin \, 2 \beta \, cos \, \theta \, (3 cos^2 \theta - 1)] cos \, \delta_3
\end{align}
The symmetry/symmetries of the resonant state/states involved in DEA can be found from the angular distribution of the fragment negative ions.\\ 
In Fig. \ref{fig:AD_All} (a), we have fitted the angular distributions of the higher energy O$^-$ ions for 9.3 and 9.9 eV electron energy on the more prominent resonant peak. As our theoretical calculation suggests, A$_1$, A$_2$, and B$_2$ resonant symmetries are present here and might be responsible for this peak; we have therefore fitted them with these resonant symmetries (dashed curves). From our fitting, we have noticed that the contribution of A$_2$ resonance is very small or negligible. Therefore, we have also fitted these distributions with the combination of  A$_1$ and B$_2$ resonant symmetries (solid curves).
The A$_1$ + B$_2$ fits justify the nature of the distributions properly.
This suggests that the higher energy O$^-$ ions are produced from the A$_1$ and B$_2$ resonances with negligible or no contribution from A$_2$ resonance. 
For lower energy ions also, we have fitted the angular distributions [Fig. \ref{fig:AD_All} (c)] with both the combination of A$_1$ + A$_2$ + B$_2$ (solid line) and A$_1$ + B$_2$ (dashed line) symmetries. From this fitting, we can see that the combination of A$_1$ + B$_2$ resonant symmetries can not justify the distribution properly, and we need to consider the contribution from the A$_2$ resonance also. 
On the other hand, the low-energy ions get highly affected by the rotation of the TNI.\cite{wrede2002quasiclassical} Therefore, the experimentally obtained angular distribution of the low-energy ions may be much different from the actual distribution. That is why it may not be a good idea to fit the angular distributions for the low energy ions using these expressions and conclude something from that.\cite{ram2012dynamics}\\
In Fig. \ref{fig:AD_All} (b), we have fitted the angular distributions of the higher energy O$^-$ ions for 10.5 and 11.1 eV electron energy on the smaller resonant peak. As our theoretical calculation suggests, A$_1$, A$_2$, and B$_2$ resonant symmetries are present here and might be responsible for this peak; we have therefore fitted them with these resonant symmetries (dashed curves). Here also, we have noticed that the contribution of A$_2$ resonance is very small or negligible. Therefore, we have also fitted these distributions with A$_1$ + B$_2$ resonant symmetries (solid curves). Here also, the A$_1$ + B$_2$ fits justify the nature of the distributions.
Therefore, the formation of higher-energy ions in this resonance may have no contribution.
We have fitted the angular distributions [Fig. \ref{fig:AD_All} (d)] with both the combination of A$_1$ + A$_2$ + B$_2$ (blue solid line) and A$_1$ + B$_2$ (black solid line) symmetries. From this fitting, we can see that the combination of A$_1$ + B$_2$ resonant symmetries can not justify the distribution properly, and we need to consider the contribution from the A$_2$ resonance also. Here also, we cannot wholly be sure about that since the axial recoil approximation gets violated most of the time for the lower energy ions.\cite{ram2012dynamics}
\begin{table*}
\small
  \caption{Fitting parameters for the angular distributions of the O$^-$ ions in the higher kinetic energy band around the bigger resonant peak.}
  \label{tbl:Table_4}
  \begin{tabular*}{\textwidth}{@{\extracolsep{\fill}}|l|l|l|l|l|}
    \hline
 Electron & Ratio of the partial waves in & Ratio of the partial waves in & Phase differences & R$^2$\\
 energy (eV) & for A$_1$ resonance ($a_0:a_1:a_2$) & for B$_2$ resonance ($b_3:b_4$) & $\delta_0$, $\delta_1$, $\delta_3$ (radian) & value\\
\hline
 &&&&\\
 9.3  & 0.40:0.39:0.88 & 0.09:0.11 & 5.20, 0.88, 0.48 &  0.93 \\
\hline
 &&&&\\ 
 9.9  & 0.48:0.19:0.90 & 0.06:0.07 & 5.72, 0.49, 1.05 &  0.96 \\  
\hline
  \end{tabular*}
\end{table*}
\begin{table*}
\small
  \caption{Fitting parameters for the angular distributions of the O$^-$ ions in the higher kinetic energy band around the smaller resonant peak.}
  \label{tbl:Table_5}
  \begin{tabular*}{\textwidth}{@{\extracolsep{\fill}}|l|l|l|l|l|}
    \hline
 Electron & Ratio of the partial waves in & Ratio of the partial waves in & Phase differences & R$^2$\\
 energy (eV) & for A$_1$ resonance ($a_0:a_1:a_2$) & for B$_2$ resonance ($b_3:b_4$) & $\delta_0$, $\delta_1$, $\delta_3$ (radian) & value\\
\hline
 &&&&\\
 10.5  & 0.46:0.27:0.94 & 0.05:0.05 & 5.44, 0.66, 1.05 &  0.98 \\
\hline
 &&&&\\ 
 11.1  & 0.38:0.42:1.00 & 0.06:0.11 & 5.07, 0.99, 0.86 &  0.98 \\  
\hline
  \end{tabular*}
\end{table*}

\section{Conclusion}
The angular distribution of the low-energy ions remains more or less similar with the increase in the incident electron energy. With the incident electron energy increase, the backward count increases, and the dip becomes prominent.
On the other hand, the angular distributions of the higher energy ions remain more or less similar with the increase in the incident electron energy. With the incident electron energy increase, the backward count decreases, and the dip near 150$^\circ$ becomes prominent. Our angular distribution agrees reasonably well with the distribution found by Gope \textit{et al.}
We have thus developed a quantitative understanding of DEA to nitrogen dioxide molecules for resonant peaks at 8.5 and 11 eV. Our experimental findings are justified by the theoretical calculations.

\section*{Acknowledgements}
A.P. sincerely appreciates the ``Council of Scientific and Industrial Research (CSIR)'' for the financial assistance. D.N. gratefully acknowledges the financial support from the ``Science and Engineering Research Board (SERB)'' under Project No. ``CRG/2019/000872.''

\bibliographystyle{h-physrev}
\bibliography{no2_ref}

\end{document}